\documentstyle[twocolumn,prl,aps,floats,graphicx]{revtex}
\begin{document}
\draft
\wideabs{
\title{Anharmonicity, vibrational instability and Boson peak in glasses}
\author{V. L. Gurevich}
\address{Solid State Physics Division, A. F. Ioffe Institute,
194021 Saint Petersburg, Russia}
\author{D. A. Parshin}
\address{Saint Petersburg State Technical University,
195251, Saint Petersburg, Russia}
\author{H. R. Schober}
\address{Institut f\"ur Festk\"orperforschung, Forschungszentrum
J\"ulich, D-52425, J\"ulich, Germany}
\date{\today}
\maketitle
\begin{abstract}
We show that a {\em vibrational instability} of the spectrum of
weakly interacting quasi-local harmonic modes creates the maximum in
the inelastic scattering intensity in glasses, the Boson peak. The
instability, limited by anharmonicity, causes a complete
reconstruction of the vibrational density of states (DOS) below some
frequency $\omega_c$, proportional to the strength of interaction.
The DOS of the new {\em harmonic modes} is independent of the actual
value of the anharmonicity. It is a universal function of frequency
depending on a single parameter -- the Boson peak frequency,
$\omega_b$ which is a function of interaction strength.  The excess
of the DOS over the Debye value is $\propto\omega^4$ at low
frequencies and linear in $\omega$ in the interval $\omega_b \ll
\omega \ll \omega_c$. Our results are in an excellent agreement with
recent experimental studies.
\end{abstract}
\pacs{PACS numbers: 61.43.Fs, 63.50+x, 78.30.Ly}
}

\section{Introduction}

One of the most striking properties of glasses is a maximum in the
inelastic scattering intensity observed in neutron and Raman
scattering at frequencies between 0.5 and 2~THz, far below the Debye
frequency \cite{phillips:81}. This so called Boson peak (BP)
indicates an excess of low frequency vibrations over the Debye value
which is given by the sound waves. It is seen as a maximum in the
vibrational DOS divided by $\omega^2$, $G(\omega) /\omega^2$ but
not necessarily in $G(\omega)$ itself~\cite{Ahmad:86}. The BP is a
general feature, its magnitude varies strongly between materials.

Despite numerous efforts, the BP remains one of the most intriguing
problems of solid state physics.  Some authors attribute the BP to
vibrations of clusters of atoms of typical sizes
\cite{malinovsky:91,pang:92}. The physical origin of these
clusters in homogeneous amorphous media remains unclear and they
have not been identified in numerical simulations.

Another popular qualitative explanation of the Boson peak is a
softening of acoustic phonons by static disorder
\cite{elliott:92,masciovecchio:99} due to elastic Rayleigh
scattering. However, even the most optimistic estimates show that the
Rayleigh contribution is at least 4 times too small to explain the
experimental data on thermal conductivity in
glasses~\cite{elliott:92}. This mechanism is also in contradiction to
the linear dispersion law for acoustical phonons at the Boson peak
frequency seen in molecular dynamics in Ref.~\cite{tarell}.

Sometimes the BP is related to low lying optic modes of parental
crystals~\cite{FSS,dove:97,SSPK}. Whereas the BP is a general feature
of glasses such crystal structures with soft optic modes cannot be
identified always.  Such a mechanism is possible in some cases. Yet
it remains a puzzle how these crystalline peaks are transformed in
the glassy state to a shoulder in the vibrational DOS.

Recent work on harmonic lattice models demonstrated that softening of
disordered force constants can smear and push to low frequencies peaks
which exist in the crystalline
DOS~\cite{schirmacher:98,taraskin:01,kantelhardt:01}. 
In another approach the vibrations of a random distribution of atoms,
interacting  with a Gaussian-shaped pair potential, was studied
\cite{grigera:01} in a harmonic scalar approximation. Reducing the density
the system becomes unstable. Approaching this instability a low
frequency peak appears in $G(\omega) /\omega^2$ which resembles the
BP.  The main drawback of these models is their neglect of the static
displacement of the atoms in response to disorder. In real glasses
where short range order is conserved such feedback always occurs
since the forces between the atoms strongly depend on their distance.
In particular, the above models have no built in mechanism to
stabilize vibrations with negative $\omega^2$, unstable modes.

The proposed models of the Boson peak do not account for
anharmonicity effects which, as we will show in the present paper,
become very important especially for small force constants. Glasses,
at low frequencies/temperatures, are highly anharmonic as seen in
most of their macroscopic thermodynamic functions.  Anharmonicity and
static displacements, together, stabilize otherwise unstable
vibrational modes.  This does, however, not imply that the vibrations
at the Boson peak are anharmonic.  Anharmonicity is essential in
forming the equilibrium structure, and thus the force constants which
determine the {\em proper harmonic spectrum} of low frequency modes.

Another important point is that the previous explanations do not
relate the Boson peak, which is one of the universal properties of
glasses, to other universal properties, such as the two-level systems
which dominate the low-temperature behavior or the plateau in the
thermal conductivity at moderate temperatures. This relationship
naturally emerges in our approach.

\section{Quasi-Localized Vibrations (QLV)}

We present a new universal mechanism for the formation of a BP in
glasses, out of an originally {\em flat} DOS. This arises from the
following 3 features: 1) quasi-local vibrations (resonant states)
(QLV) with a smooth, structureless initial DOS, $g_0(\omega)$ at low
$\omega$, 2) an elastic interaction between them and 3) stabilization
by anharmonicity when the system becomes unstable because of
interaction.

As the well known two level systems, the QLV are a typical feature of
disordered systems. They are additional modes and are characterized
by a large vibrational amplitude of some group of atoms. Their
existence in glasses was predicted in Refs.~\cite{KP1,MAK,UB}. They
can be described as low frequency harmonic oscillators (HO) which
couple bilinearly to the sound waves, see
Refs.~\cite{MMWI,DZ,BGGPRS}. This in turn leads to a dipole-dipole
interaction between different HO. The importance of the elastic
interaction between local defects in glasses has been stressed by Yu
and Leggett~\cite{yu:88} and Grannan {\it et al.}~\cite{grannan:90}.

The microscopic origin of QLV in disordered systems varies and
depends on the type of disorder. Broadly speaking, the QLV can be
divided into two groups. First there are materials where QLV exist
independently of the structural disorder typical for structural
glasses and amorphous solids. Orientationally disordered
(``plastic'') crystals belong to this
group~\cite{RVB,lynden:94,plastic:78}. In these materials some
molecular groups librate with low frequencies. In harmonic
approximation these soft librations can be identified with QLV. The
local potentials for the librational motion can vary from site to
site. In this case we have a distribution of the librational
frequencies. The librations couple to the sound waves which in turn
induces an interaction between them~\cite{grannan:90}. It depends on
details of the material whether the interaction is strong enough to
reconstruct the original spectrum of QLV completely or only
partially. This effect is seen in recent measurements of dielectrtric
loss spectra of ortho-carborane~\cite{LBWL}. Similarly coordination
defects in covalent materials can lead to QLV as was observed by
Biswas et al.~\cite{BBKGS} in a simulation of amorphous Si (see also
recent works~\cite{fab:97} and~\cite{feld:02}).  QLV can originate
from numerous defects such as off center ions or interstitial atoms.
Depending on the ``size'' of the defect the QLV involve more or fewer
atoms. Interstitial atoms are the prototype of a topological point
defect. QLV of interstitial atoms in FCC metals were studied
extensively in the past, see Ref.~\cite{dederichs:78} for a review.
These QLV have effective masses of four atomic masses and the crystal
structure is strongly distorted by the defect and the low frequency
of the librational QLV can be traced to the local strain. Already low
concentrations of these interstitials are sufficient to destroy the
crystalline structure completely. This is utilized in the
interstitialcy model of glass formation \cite{granato:92}.

This leads directly to the second group where the QLV result directly
from disorder. Such modes are regularly found in computer
simulations, e.g. for soft spheres~\cite{LS:91},
SiO$_2$~\cite{jin:93}, Se~\cite{OS:93}, Ni-Zr~\cite{hafner:94}, Pd-Si
and Au-Si~\cite{ballone:95}, NiB~\cite{ee:96}, in amorphous
ice~\cite{cho:94}, in amorphous and quasicrystalline
Al-Zn-Mg~\cite{hafner:93} and in simple dense fluids~\cite{wu:00}.

In these simulations the QLV were observed as localized vibrations
with frequencies below the minimal sound wave frequency allowed by
the size of the simulated sample. A simple indicator of these
``size-localized'' QLV is the scaling of the participation ratio with
system size. Increasing the system size the minimal sound wave
frequency drops and the QLV are no longer localized in the simulation
but show the typical properties of resonant modes, i.e. of low
frequency local vibrations which couple bilinearly to the sound
waves. The exact eigenvectors of the interacting system of QLV and
sound waves are superpositions of these two types of modes. The
effect of system size on the appearance of QLV in simulations was
discussed in detail in Ref.~\cite{SO:96}. There it was shown that the
exact eigenvectors at frequencies up to and above the boson peak can
be decomposed into extended sound wave like modes and the local cores
of QLV. The latter correspond to the harmonic oscillators of this work.

The physical origin of these disorder induced QLV can be traced to
local irregularities of the amorphous structure. In dense packed
metallic glasses these originate e.g. from the conflict of the local
dense packing (icosahedral packing) and global dense packing (fcc or
hcp)~\cite{Luch:00}. These local irregularities can be seen as
centers of local strains~\cite{Chen:88}.

We expect such local strain centers to be ubiquitous. The strains
will have broad distributions, which will lead to broad distributions
of QLV frequencies. Whereas local strains and QLV will be a general
property of glasses the atomistic structure of QLV reflects the
structure of the considered material. In dense packed metallic systems
the cores of the QLV have been found to be
chain-like~\cite{SOL:93,SO:96}. In SiO$_2$ they are formed by a
coupled rotations of SiO$_4$ tetrahedra~\cite{UB,Olig:99}. In Se one
has coupled chains and rings~\cite{OS:93}, etc.

Another possible (and natural) mechanism of the QLV formation are low
lying optical modes in parental crystals. Disorder in amorphous
material would destroy the long range coherence of optical modes.
This makes them practically indistinguishable from quasi-local modes.

Together with the tunneling systems, the QLV
form the main ingredient of the {\it soft potential
model}~\cite{KKI,IKP} (see review~\cite{DPR}). They manifest
themselves in experimental values, e. g.  the excess specific
heat~\cite{IKP,BGGS,PLBL,DCT} and the plateau in the thermal
conductivity~\cite{BGGPRS,RB}, in inelastic light~\cite{GPPS} and neutron
scattering~\cite{UB} and they are observed in many numerical
calculations, cited above.

\section{Density of states of Quasi-Localized Vibrations (QLV)}

One might think {\em a priori} that the QLV can have an arbitrary
DOS, $g(\omega)$, depending on the particulars of the glass. We will
show, that due to the interaction between the HO, $g(\omega)$ is a
{\em universal} function at low frequencies.  This universality stems
from the {\it vibrational instability} of the spectrum which occurs
in nearly all systems of interacting HO.  Anharmonicity stabilizes
the system in new minima, and thus reconstructs the DOS to a new {\em
harmonic} spectrum. As a result $g(\omega)/\omega^2$ acquires a
maximum without a peak in $g(\omega)$ itself.

%%%%%%%%% Picture 1   %%%%%%%%%%%%%%%%
\begin{figure}[htb]
%\unitlength=1cm
%\begin{picture}(0,6.5)
\includegraphics[bb=360 580 600 770,angle=-180,totalheight=6cm,keepaspectratio]{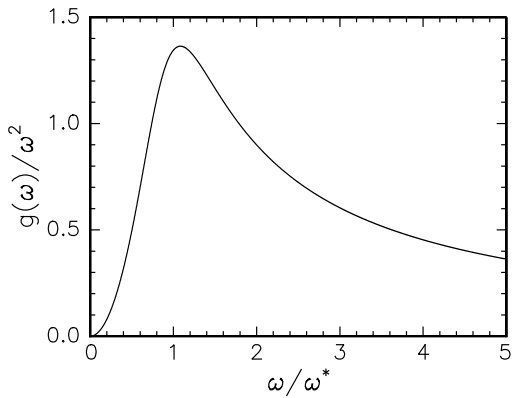}
%\put(0.5,6.3){\special{em:graph bp_fig1.pcx}}
\caption{The Boson peak, Eq.~(\protect\ref{bp3}).}
\label{fig:bpeak}
%\end{picture}
\end{figure}
%%%%%%%%% End Picture 1 %%%%%%%%%%%
Below we will derive the following form of the reduced DOS
of these harmonic resonant modes (excluding the Debye part at low
$\omega$ which is not seen in Raman scattering)
\begin{eqnarray}
\label{bp3}
{g(\omega)\over\omega^2}={3C\over\pi\omega^\star}
\left({\omega^\star\over\omega}\right)^4\left[z_1^2(\omega)
+z_2^2(\omega)\right]^{-1}\nonumber\\
\times\left[\frac{1}{2z_1(\omega)}\ln\frac{z_1(\omega)
+1}{z_1(\omega)-1} +
\frac{1}{z_2(\omega)}\tan^{-1}\frac{1}{z_2(\omega)} \right]
\end{eqnarray}
where $C$ is a constant (Eq.~(\ref{bp1})) and
\begin{equation}
\label{bp2}
z_{1,2}(\omega)={1\over2}\sqrt{\sqrt{9+8(\omega^\star/\omega)^6}\pm3}.
\end{equation}

The function $g(\omega)/\omega^2$ is plotted in Fig.~\ref{fig:bpeak}.
It depends on a single parameter, $\omega^\star$ characterizing the
position of the Boson peak. The maximum of $g(\omega)/\omega^2$, the
Boson peak, is at $\omega_b\approx1.1\omega^*$. For small
frequencies, $\omega\ll\omega_b$,\, $g(\omega)\propto\omega^4$ while
for large ones, $\omega\gg\omega_b$,\, $g(\omega)\propto\omega$.

\section{Vibrational instability}

To illustrate our central idea of a vibrational instability, we start
with a pair of interacting HO immersed in an elastic continuum.
The potential energy is given by
\begin{equation}
U_{\rm har}(x_1,x_2)= M_1\omega_1^2x_1^2/2 +
M_2\omega_2^2x_2^2/2 - I_{12}x_1x_2.
\label{2osc}
\end{equation}
Here $x_{1,2}$ are the HO coordinates, $M_{1,2}$ the masses and
$\omega_{1,2}$ the bare frequencies of the two HO, i. e. neglecting
the bilinear interaction. The interaction strength is given
by~\cite{GPPS}
\begin{equation}
I_{12}=g_{12}J/r^3_{12},\quad J\equiv \Lambda^2/\rho v^2
\label{hto1}
\end{equation}
where $g_{12}$ accounts for the relative orientation of the HO,
$r_{12}$ is their distance, $\rho$ is the mass density of the glass
and $v$ is a sound velocity. The interaction between the HO is due
to the coupling between a single HO and the surrounding elastic
medium (the glass). This HO-phonon
coupling has the form~\cite{BGGPRS}
$$
{\cal H}_{\rm int} = \Lambda x\varepsilon ,
$$
where $\Lambda$ is the coupling constant and $\varepsilon$ the
strain.

Diagonalization of Eq.~(\ref{2osc}) yields two frequencies
\begin{equation}
\widetilde{\omega}_{1,2}^2=\frac{\omega_1^2+\omega_2^2}{2} \mp
\sqrt{\left( \frac{\omega_1^2-\omega_2^2}{2}\right)^2 +
\frac{I_{12}^2}{M_1M_2}}.
\label{osc2}
\end{equation}
The smaller value, $\widetilde{\omega}_1^2$, becomes negative
when the interaction $I_{12}\equiv I$ exceeds the threshold
(critical) value 
\begin{equation}
I_c\equiv\omega_1\omega_2\sqrt{M_1M_2}.
\label{eq:ytf1}
\end{equation}

This instability persists also in a system of many interacting HO.
In a real physical system, anharmonic forces always stabilize an
embedded HO in a nearby minimum of the potential energy. The position
of this minimum depends on the interaction between HO. We are thus
confronted with the many-body problem of finding the minima of the
potential energy for a system of interacting anharmonic oscillators,
similar to the one considered in Ref.~\cite{grannan:90,KH}. The new
frequencies, in these new minima, are real and different from the
original ones. The harmonic vibrational spectrum is reconstructed. We
will call this {\em anharmonicity limited vibrational instability}.

\section{Stabilization by anharmonicity}

We will now show that for {\em weak} interaction, $I$, the
reconstructed DOS has, below a characteristic frequency
$\omega_c\propto |I|$,a universal form irrespective of its original
form. First, due to interaction, it becomes a {\em linear} function
of frequency, $g(\omega)\propto\omega$.  Secondly, the displacements
of the previously unstable oscillators from their old equilibrium
positions create static random forces which cause a {\em second}
reconstruction of the DOS below another frequency
$\omega_b\ll\omega_c$. Due to so called {\em sea-gull
singularity}~\cite{IKP} at $\omega = 0$ the linear DOS is
reconstructed to $g(\omega)\propto\omega^4$ for $\omega << \omega_b$.
Together, these two reconstructions produce a maximum of
$g(\omega)/\omega^2$ at $\omega = \omega_b$.

Let us consider a number of randomly distributed, interacting HO with
concentration $n_0$ and an initial DOS,
$g_0(\omega)$  (normalized to unity), in the frequency range from 0 to
$\omega_0$, where $g_0(\omega)$ is a monotonously increasing function
of $\omega$. For the harmonic part of the interaction we take the
generalization of Eq.~(\ref{2osc}) and add an anharmonic term to
stabilize the system
\begin{equation}
U_{\rm anhar}=(1/4)\sum_i A_ix_i^4, \quad A_i>0 .
\label{anh}
\end{equation}

We will take the interaction $I$ to be the {\em small parameter} of
our theory, i.e. we assume that the typical random interaction, $I$,
between neighboring HO is much smaller than the typical
values of $M\omega_0^2$. As $|I| \ll M\omega_0^2$
frequencies of order $\omega_0$ will be practically unaffected
by the interaction whereas HO with frequencies $\omega < \omega_c$
will be displaced to new minima,
where
\begin{equation}
\omega_c\simeq
|I|/M\omega_0 \ll \omega_0.
\label{omc}
\end{equation}

Since the concentration of unstable HO is much smaller than the one
of the stable ones a low frequency oscillator is typically surrounded
by high frequency ones. We can simplify our consideration by again
considering pairs of HO, one with a low frequency
$\omega_1\lesssim\omega_c$ and the other one from the bulk of the
HO with a frequency $\omega_2$ of the order of $\omega_0$ (see
Appendix~\ref{sec:cluster} for the general case).
Due to the combined action of interaction and anharmonicity the two
HO will be displaced into new minima, $x_{10}$ and $x_{20}$,
given by the equations ($I_{12}\equiv I$)
\begin{equation}
\begin{array}{rcl}
Ix_{20}&=&x_{10}(M_1\omega_1^2+A_1x_{10}^2) ,\\[2pt]
Ix_{10}&=&x_{20}(M_2\omega_2^2+A_2x_{20}^2) .\\
\end{array}
\label{pos1}
\end{equation}

For $|I|>I_c$ we need the nonzero solutions of these
equations. Expanding around either minimum we find
the new (harmonic) frequencies from the secular equation
\begin{equation}
\left|
\begin{array}{cc}
\alpha_1-M_1\omega^2&-I\\
-I&\alpha_2-M_2\omega^2\\
\end{array}
\right| = 0
\label{deter1}
\end{equation}
with
\begin{equation}
\alpha_i=M_i\omega_i^2+3A_ix_{i0}^2 , \quad i=1,2.
\label{alphai}
\end{equation}

>From the condition $\omega_1\ll \omega_2$ follows $x_{20}\ll
x_{10}$ and, therefore, the term $A_2x^2_{20}$ in Eq.~(\ref{pos1})
can be neglected giving
\begin{equation}
x_{20} = (I/M_2\omega^2_2)\,x_{10} .
\label{gyt4}
\end{equation}
and
\begin{equation}
x_{10}=\omega_1\sqrt{M_1/A_1}\,\sqrt{(I/I_c)^2 -1} .
\label{iop7}
\end{equation}

As a result we get from (\ref{deter1}) under the condition
$\omega_1/\omega_2\ll 1$ with this accuracy the new frequencies
$\widetilde{\omega}_2=\omega_2$ and
\begin{equation}
\widetilde{\omega}_1^2=2\omega_1^2\left[(I/I_c)^2-1\right] .
\label{new1}
\end{equation}
The smaller frequency (\ref{new1}) is the solution of the linear
equation (compare with Eq.~(\ref{deter1}))
\begin{equation}
M_2\omega_2^2(\alpha_1-M_1\omega^2)=I^2.
\label{lin22}
\end{equation}
It is remarkable that for weak interaction the strength of the
anharmonicity $A_i$ does not enter the renormalized frequency
(\ref{new1}).

Near the threshold where $(|I|-I_c)/I_c\ll 1$, the smaller frequency
squared $\widetilde{\omega}_1^2$ is proportional to $(|I|-I_c)/I_c$.
Provided the distribution of the random quantity $I$ is smooth one
gets, therefore, below $\omega_c$ a linear DOS
($\widetilde{g}(\omega)\propto\omega$) irrespective to the initial form of
$g_0(\omega)$. In Appendix \ref{sec:cluster} it is shown that the same
result holds if one has a low frequency HO surrounded by several high
frequency ones. Our numerical calculations (see Section \ref{sec:num})
also show that this case is typical.

\section{The Boson peak}

If the low-frequency HO with their reconstructed linear DOS
were isolated, the problem would be solved.  There is, however,
a further interaction between these oscillators which we have not
taken into account so far. The low-frequency HO, displaced from their
equilibrium positions, create random static forces $f$. The force
$f_i$ exerted on the $i$th oscillator by the $j$th one is
\begin{equation}
f_i=I_{ij}x_{j0}.
\label{f}
\end{equation}

In a purely harmonic case, these linear forces would not affect the
frequencies. Anharmonicity, however, renormalizes the low frequency
part of the spectrum~\cite{IKP}, a manifestation of the {\em sea-gull
singularity}. Consider an anharmonic oscillator under the action of a
random static force $f$
\begin{equation}
U(x)=Ax^4/4 + M\omega_1^2x^2/2 - fx
\label{ao1}
\end{equation}
where $\omega_1$ is the oscillator frequency in the harmonic
approximation. The force $f$ shifts the equilibrium position from
$x=0$ to $x_0\neq 0$, given by
\begin{equation}
\label{xce8}
Ax_0^3 + M\omega_1^2x_0 - f =0 ,
\end{equation}
where the oscillator has a new (harmonic) frequency
\begin{equation}
\label{kiu6}
\omega^2_{\rm new} = \omega_1^2+3Ax_0^2/M.
\end{equation}
If $\widetilde{g}_1(\omega_1)$ is the distribution function of frequencies
$\omega_1$ and $P(f)$ is the distribution of random forces, then the
renormalized DOS is given by
\begin{equation}
\label{des2}
g(\omega)=\int\limits_0^\infty \widetilde{g}_1(\omega_1)d\omega_1
\int\limits_{-\infty}^{\infty}dfP(f)\delta\left(\omega -
\omega_{\rm new}\right) .
\end{equation}

As the forces between the HO are proportional to
$r_{ij}^{-3}$ their sum is Lorentzian distributed (see Appendix
\ref{sec:holts}):
\begin{equation}
\label{tik1}
P(f)=\frac{1}{\pi}\frac{\delta f}{f^2+(\delta f)^2}.
\end{equation}
Assuming $\omega\ll\omega_c$ and integrating Eq.~(\ref{des2})
with $\widetilde{g}_1(\omega_1)=C\omega_1$ we arrive at the integral
\begin{equation}
\label{bp1}
{g(\omega)\over\omega^2} =  {6C\over\pi\omega^\star}
\left({\omega\over\omega^\star}\right)^2\int\limits_0^1
\frac{\displaystyle dt}{\displaystyle
1+\left(\omega/\omega^\star\right)^6t^2(3-2t^2)}
\end{equation}
with
\begin{equation}
\label{fre1}
\omega^\star = {\sqrt{3}A^{1/6} (\delta f)^{1/3}/\sqrt{M}}
\end{equation}
and after integration finally Eq.~(\ref{bp3}) is obtained.

For small frequencies, below the Boson peak, $\omega\ll\omega_b$
only small random forces $f$ contribute to the second integral in
Eq.~(\ref{des2}). In this case the distribution function $P(f)$ can be
approximated by a constant value, $P(0)$, and we get from (\ref{des2})
\begin{equation}
g(\omega)\propto \omega^3 \int\limits_0^\omega d\omega_1
\frac{\omega_1}{\sqrt{\omega^2-\omega_1^2}}\propto \omega^4 .
\label{whh}
\end{equation}

As a result at low frequencies the renormalized DOS,
$g(\omega)\propto \omega^4$, Ref.~\cite{IKP}. For sufficiently large
frequencies, $\omega\gg\omega_b$ the action of random static forces
on the HO spectrum can be discarded. In this case we recover the
linear DOS, $g(\omega)\propto\omega$.

The BP frequency, $\omega_b \approx 1.1 \omega^\star$, is determined
by the characteristic value of the random static force $\delta f$,
Eq.~(\ref{f}), acting on an HO with the characteristic
frequency $\omega_c$. According to Eq.~(\ref{f}), it is due to the
interaction between HO with frequencies of order of
$\omega_c$, i.e.
$$
I_{ij}^{(c)}\approx Jn_c, \quad J\approx I/n_0\approx
M\omega_c/n_0g_0(\omega_0)
$$
where $n_c\approx n_0g_0(\omega_c)\omega_c$ is
the concentration of these HO. The characteristic
displacement of a low-frequency HO (\ref{iop7}) from the equilibrium
position is $x_{j0}\approx \omega_c\sqrt{M/A}$.  As a result, we get
the estimate
\begin{equation}
\delta f\approx M\sqrt{\frac{M}{A}}\,\omega_c^3\,
\frac{g_0(\omega_c)}{g_0(\omega_0)}
\label{eq:gh2}
\end{equation}
and according to (\ref{fre1})
\begin{equation}
\omega_b \approx \omega_c \left[g_0(\omega_c)/g_0(\omega_0)
\right]^{1/3},\quad \omega_b \ll \omega_c  .
\label{k}
\end{equation}
Again, in lowest order the anharmonicity $A$ does not enter this
formula.

As a result we get a following estimate for the
reconstructed DOS
\begin{equation}
g(\omega) \simeq \left\{ 
\begin{array}{ll}
g_0(\omega), & \omega>\omega_c, \\
\omega\, g_0(\omega_c)/\omega_c, & \omega_b<\omega<\omega_c,\\
\omega^4g_0(\omega_0)/\omega_c^4 , & \omega<\omega_b .
\end{array} \right.
\label{eq:est1}
\end{equation}  

If the DOS of the noninteracting oscillators is given by  a power law, 
$g_0(\omega) \propto \omega^n$, the BP frequency $\omega_b$ scales
with the interaction strength $I$ as
\begin{equation}
\omega_b\propto |I|^{1+n/3} .
\label{scal1}
\end{equation}
Since in accordance with (\ref{omc}) $\omega_c\propto |I|$, we have
always $\omega_b\ll\omega_c$ for $n>0$.

\section{Numerical simulation}
\label{sec:num}

%%%%%%%%% Picture 2   %%%%%%%%%%%%%%%%
\begin{figure}[htb]
\includegraphics[bb=360 580 600 770,angle=-180,totalheight=6cm,keepaspectratio]{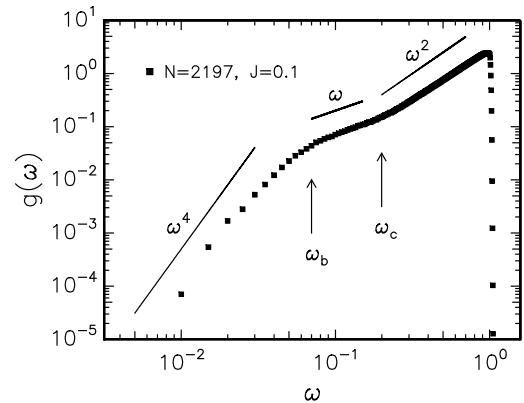}
%\unitlength=1cm
%\begin{picture}(0,7.2)
%\put(0.5,7.2){\special{em:graph bp_fig2.pcx}}
\caption{Simulated density of states ($g_0(\omega)\propto \omega^2$,
$N=2097$) in a log-log representation. The arrows indicate the two
characteristic frequencies $\omega_b$ and $\omega_c$.}
\label{fig:dos1}
%\end{picture}
\end{figure}
%%%%%%%%% End Picture 2 %%%%%%%%%%%
To test our ideas by numerical simulations, we placed $N$
oscillators with frequencies $0<\omega_i<1$ on a simple cubic lattice
with lattice constant $a=1$ and periodic boundary conditions. To
simulate random orientations of the oscillators we took for $g_{ij}$,
Eq.~(\ref{hto1}), random numbers in the interval $[-0.5,0.5]$.  The
masses, $M_i$, and anharmonicity parameters, $A_i$, were put to 1.  The
DOS for the noninteracting oscillators was taken as $g_0(\omega)
\propto \omega^n$, with
$n=1,2,3$.

Using the potential energy given by the generalization of
Eq.~(\ref{2osc}) plus the anharmonicity (Eq.~(\ref{anh})) we then
minimized the potential energy, and in the usual harmonic expansion
around this minimum calculated the DOS for different interaction
strengths, $J$. This was repeated for up to 10000 representations.
To check for size dependence we did the calculations for different
$N$. Apart from the case $J=0.07$ the results did not change between
$N=2097$ and $N=4096$.

The predicted change-over in the $\omega$-dependence of the DOS at
two characteristic frequencies $\omega_c$ and $\omega_b$ and the linear
part in between can be clearly observed in a log-log representation,
Fig.~\ref{fig:dos1}, for $g_0(\omega)\propto\omega^2$ and $J=0.1$.
>From the calculated eigenvectors we find that, as expected, at the
lowest frequencies the HO are weakly coupled whereas near and above
$\omega_b$ the eigenmodes are complicated superpositions of many HO.

%%%%%%%%% Picture 3   %%%%%%%%%%%%%%%%
\begin{figure}[htb]
\includegraphics[bb=360 580 600 770,angle=-180,totalheight=6cm,keepaspectratio]{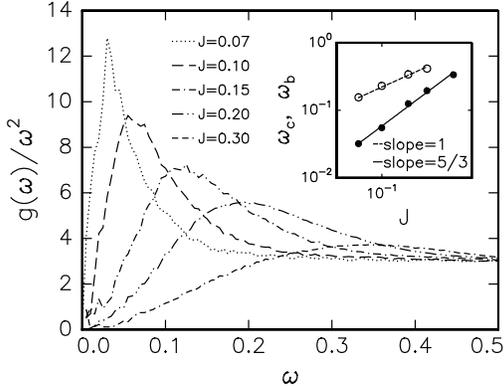}
%\unitlength=1cm
%\begin{picture}(0,7.5)
%\put(0.5,7.5){\special{em:graph bp_fig3.pcx}}
\caption{Simulated $g(\omega)/\omega^2$ for
different interaction strengths ($g_0(\omega)\propto \omega^2$,
$N=2097$ and $N=4096$ ($J=0.07$)). The insert shows the scaling of the
crossover frequencies $\omega_c$ ($\protect\circ$) and
$\omega_b$ ($\protect\bullet$) with interaction strength $J$.}
\label{fig:bosp2}
%\end{picture}
\end{figure}
%%%%%%%%%   End Picture 3   %%%%%%%%%%%
Fig.~\ref{fig:bosp2} shows the dependence of the simulated
$g(\omega)/\omega^2$ on the interaction strength $J$. We can see the
general increase of $\omega_b$  and related decrease of the BP intensity
with increasing $J$. Our simulations cover one decade in
BP frequencies. The insert shows that, in full agreement with our
predictions (see Eq.~(\ref{omc}) and Eq.~(\ref{scal1})), the crossover
frequencies change with interaction $J$ as $\omega_c\propto J$ and
$\omega_b \propto J^{1+n/3}$.

\section{Discussion and comparison with experiment}
\label{sec:disc}

%%%%     Picture 4  %%%%%%%%%%%%%
\begin{figure}[hbt]
\includegraphics[bb=360 580 600 770,angle=-180,totalheight=6cm,keepaspectratio]{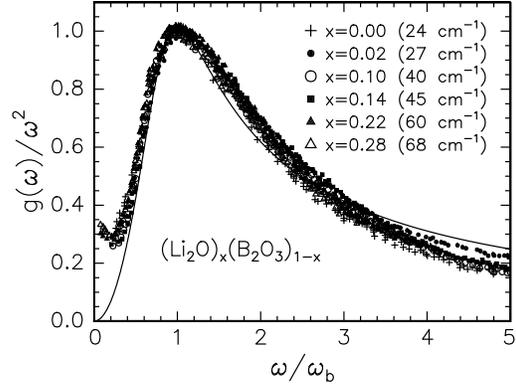}
%\unitlength=1cm
%\begin{picture}(0,7.5)
%\put(0.5,7.5){\special{em:graph bp_fig4.pcx}}
\caption{Boson peak in reduced units: Eq.~(\protect\ref{bp3})
(solid line) and Raman data for lithium borate glasses~\protect\cite{LBG}.
The positions of the Boson peak (for different compositions
$x$) are given in brackets.}
\label{fig:boson0}
%\end{picture}
\end{figure}
%%%% End Picture 4  %%%%%%%%%%%%%%
In Fig.~\ref{fig:boson0} we compare our theoretical curve,
Eq.~(\ref{bp3}), with Raman scattering data of lithium borate
glasses~\cite{LBG} with different compositions. The agreement is
remarkably good over the whole composition range. This supports the
idea of a universal shape of the Boson peak~\cite{MS}. The shift of
the BP to higher frequencies with increasing concentration of Li$_2$O
can be explained by an increase of the total concentration of QLV and
consequently of their interaction.

%%%%%%%%% Picture 5   %%%%%%%%%%%%%%%%
\begin{figure}[htb]
\includegraphics[bb=360 580 600 770,angle=-180,totalheight=6cm,keepaspectratio]{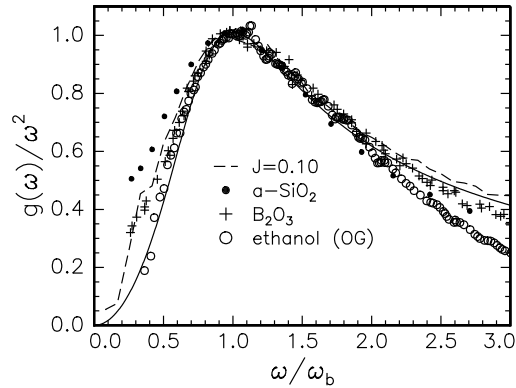}
%\unitlength=1cm
%\begin{picture}(0,7.5)
%\put(0.5,7.5){\special{em:graph bp_fig5.pcx}}
\caption{Boson peak in reduced units: Eq.~(\protect\ref{bp3}) (solid line),
numerical simulation, Fig.~3, (dashed line), neutron scattering data for
a-SiO$_2$ at $T=51 K$~\protect\cite{WB} ($\bullet$) and for the
orientational glass phase of ethanol~\protect\cite{RVB} ($\circ$) and
Raman data for a-B$_2$O$_3$~\protect\cite{LBG} (+).}
\label{fig:boson}
%\end{picture}
\end{figure}
%%%%%%%%% End Picture 5 %%%%%%%%%%%
The agreement between theory and experiment is not confined to this
class of material. This is exemplified by Fig.~\ref{fig:boson} which
shows a comparison of the theoretical curve with numerical simulation
results and neutron and Raman scattering data for 3 different glasses.

\begin{figure}[htb]
\includegraphics[bb=360 580 600 770,angle=-180,totalheight=6cm,keepaspectratio]{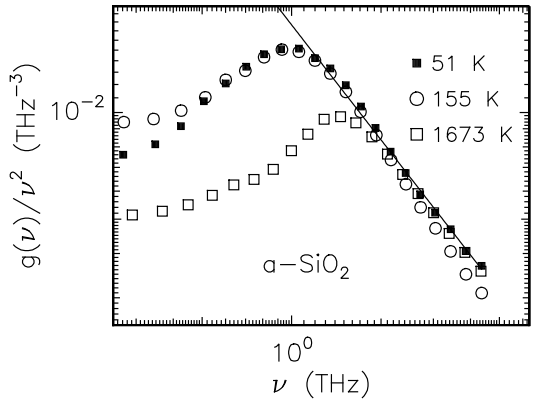}
%\unitlength=1cm
%\begin{picture}(0,7.5)
%\put(0.5,7.5){\special{em:graph bp_fig6.pcx}}
\caption{Density of states of vitreous silica for three temperatures
taken from inelastic neutron scattering data~\protect\cite{WB}. The
slope of the straight line on the figure is equal to -1.1.}
\label{fig:buch_bp}
%\end{picture}
\end{figure}

One of the most important results of our theory is the predicted linear
frequency dependence of the density of vibrational states above the
Boson peak.  It stems from the vibrational instability of interacting
harmonic modes. Such linear behavior has been observed in many
numerical simulations on different glasses and model disordered
systems~\cite{LS:91,hafner:94,ballone:95,hafner:93,OS:97,meshk:97}.
It is also in a good agreement with many experimental
results~\cite{inamura:99,WB,cappel:95,carp:85,phill:89,berm:93,ahm:93,suck:01}
where the vibrational DOS has a section with near linear frequency
dependence. Fig.~\ref{fig:buch_bp} shows this for vitreous silica at
different temperatures. Above the Boson peak the DOS increases
approximately linearly with frequency.
%%%%%%%%% Picture  6  %%%%%%%%%%%%%%%%
%%%%%%%%% End Picture  6   %%%%%%%%%%%

In this paper we dealt with the case of weak interaction between HO.
If the interaction is increased the characteristic frequencies
$\omega_b$ and $\omega_c$ grow and the gap between them narrows and
finally disappears. Then our "BP" in $g(\omega)/\omega^2$
superimposes the ``boundary peak'' in $g_0(\omega)$ at the edge of
the assumed spectrum of naked (noninteracting) QLV.  The BP can no
longer be distinguished from the boundary peak or from a possible
equivalent maximum in $g_0(\omega)$.  It is possible that in some
cases, e.g. orientational glasses, $g_0(\omega)$ has a pronounced
peak which is still visible after the reshaping of the DOS by
interaction.

Similar models with strong coupling between oscillators were
previously investigated by molecular dynamic
simulations~\cite{grannan:90} and by the replica method~\cite{KH}.
For example in the Ref.~\cite{grannan:90} Grannan {\it et al.}
assumed that the dynamics "is completely dominated by the interaction
between defects mediated by the strain field". The local field
(harmonic and anharmonic contribution) was neglected. We consider the
opposite case where the local field (harmonic and anharmonic)
dominates and interaction between dipoles is weak.

\section{Conclusion}
\label{sec:concl}

In conclusion we present a universal picture of the BP formation in
glasses. We have shown that the low frequency quasi-localized
harmonic modes in glasses are destabilized by the weak bilinear
interaction between them. Anharmonicity stabilizes the system in a
new minimum of configuration space. It completely reconstructs the
low frequency part of the spectrum (at $\omega<\omega_c$) and the
Boson peak feature (at $\omega_b\ll\omega_c$) naturally emerges.  The
thus created boson peak has a material independent shape.  At low
frequencies, below the BP, the vibrational DOS increases as
$g(\omega)\propto\omega^4$, and above the BP, as
$g(\omega)\propto\omega$.

Although the anharmonicity is responsible for this effect, the final
spectrum of stable vibrations remains {\em harmonic}.  A remarkable
feature of the presented theory is that the strength of the
anharmonicity does not enter the new stable spectrum at all. It looks
as if the anharmonicity does all the work, it stabilizes the system
in a new minimum and reconstructs the spectrum and then, like a
ghost, disappears. Therefore, the discussed phenomenon is independent
of the variation of the anharmonicity between different materials.
The only parameter entering the final density of states is the
strength of interaction between the HO.

Compared to previous work, a main new result of our approach is the
natural emergence of the BP on the unstructured, flat low frequency
part of the initial spectrum $g_0(\omega)/\omega^2$ where the DOS
previously had no peaks.  For small interactions the BP frequency is
much smaller than the Debye frequency value. It shifts with
interaction strength, $I$, which explains the large variety of BP
magnitudes found in experiment. In contrast with previous models, a
natural connection of the Boson peak phenomena with other universal
properties of glasses is established.

The authors VLG and DAP gratefully acknowledge the financial support
of the German Ministry of Science and Technology and the hospitality
of the Forschungszentrum J\"ulich where part of the work was done.

\appendix

\section{Cluster approach}
\label{sec:cluster}

Consider a cluster containing a low frequency
oscillator with frequency $\omega_1\lesssim\omega_c$
surrounded by a large number, $s-1$, of HO with much higher frequencies
$\omega_j\sim\omega_0$. Inclusive their interaction, the total
potential energy of the cluster is
\begin{equation}
U_{\rm tot}=\sum_i\frac{M_i\omega_i^2}{2}+
\frac{1}{4}\sum_i A_ix_i^4 -
\frac{1}{2}\sum_{i,j\ne i}I_{ij} x_ix_j .
\label{potgen}
\end{equation}
The equilibrium positions of the HO, $x_{i0}$, are given by the
system of $s$ nonlinear equations
\begin{equation}
M_i\omega_i^2x_{i0}+A_ix_{i0}^3=\sum_{j\ne i}I_{ij}x_{j0},
\quad i=1,2,...,s.
\label{neweq}
\end{equation}

In the case of instability ($x_{i0}\ne 0$), in analogy to the
previously considered case of a pair of oscillators, the static
displacements of the high-frequency oscillators are much smaller than
the one of the low-frequency oscillator, $x_{10}$. Therefore, in
leading order 
\begin{equation}
x_{i0}=(I_{1i}/M_i\omega_i^2)x_{10}, \quad i\ne 1 .
\label{eq:stat1}
\end{equation}
Inserting these values into Eq.~(\ref{neweq}) for $i=1$ we get
\begin{equation}
M_1\omega_1^2x_{10}+A_1x_{10}^3 =
x_{10}\sum_{i\ne 1} \frac{I_{1i}^2}{M_i\omega_i^2} .
\label{x10}
\end{equation}

Under the condition
\begin{equation}
M_1\omega_1^2 < k \quad \mbox{where} \quad
k\equiv\sum_{i\ne 1}\frac{I_{1i}^2}{M_i\omega_i^2} 
\label{crit1}
\end{equation}
the cluster becomes unstable and the low frequency oscillator is
displaced to a new minimum
\begin{equation}
x_{10}=\sqrt{(k-M_1\omega_1^2)/A_1} .
\label{equl1}
\end{equation}
In the opposite case, $M_1\omega_1^2>k$, the cluster is stable and
$x_{i0}=0$.

The eigenfrequencies of the interacting oscillators are the solutions
of the secular equation of order of $s$
\begin{equation}
\left|
\begin{array}{cccc}
\alpha_1-M_1\omega^2 & -I_{12}&\ldots & -I_{1s}\\
-I_{21}& \alpha_2-M_2\omega^2 & \ldots & -I_{2s}\\
\vdots &\vdots & \ddots &\vdots \\
-I_{s1} & -I_{s2}& \ldots & \alpha_s-M_s\omega^2
\end{array}
\right| = 0 .
\label{seqeq2}
\end{equation}
Here the $\alpha_i$ are given by Eq.~(\ref{alphai}). In leading order
in $I_{ij}/M\omega_0^2$ the secular equation is a linear
equation for $\omega^2$ (compare Eq.~(\ref{lin22}))
\begin{equation}
(\alpha_1-M_1\omega^2)\prod_{j\ne 1}M_j\omega_j^2 -
\sum_{i\ne 1}I_{1i}^2\prod_{j\ne 1,i}M_j\omega_j^2 = 0
\label{det12}
\end{equation}
or
\begin{equation}
\alpha_1-M_1\omega^2 = \sum_{i\ne 1}\frac{I_{1i}^2}{M_i\omega_i^2}
= k
\label{lineq2}
\end{equation}
and the new low frequency of the system of coupled oscillators is
given by 
\begin{equation}
\widetilde{\omega_1}^2 =\left\{{\displaystyle{1\over M_1}(M_1
\omega_1^2-k), \quad k<M_1\omega_1^2\atop
\displaystyle{2\over M_1}(k-M_1\omega_1^2),\quad
k>M_1\omega_1^2\,.}\right.
\label{1}
\end{equation}
As in the case of a pair of oscillators, the anharmonicity has been used
in the derivation of Eq.~(\ref{1}) but does not appear in this
or our final result, Eq.~(\ref{bp3}).

To derive of the new (reconstructed) DOS the distribution of $k$,
$\rho(k)$, has to be calculated.  Inserting Eqs.~(\ref{hto1}) and
(\ref{crit1}) into the definition of $\rho(k)$ gives
\begin{equation}
\rho(k)=\left\langle\delta\left(k-\frac{J^2}{M}\sum_{j\ne 1}\frac{g_{1j}^2}
{r_{1j}^6 \omega_j^2}\right)\right\rangle.
\label{crit3}
\end{equation}
Here the angular brackets denote averaging over the positions of
the $s-1$ high frequency HO, their frequencies and orientations. For
simplicity we take equal masses $M_j=M$ and for $g_{ij}$ a uniform
distribution in the interval [-1/2, 1/2].

Using the Holtsmark method~\cite{H} (see Appendix \ref{sec:holts})
one gets
\begin{equation}
\rho(k)=\frac{1}{\sqrt{2\pi}}\frac{B}{k^{3/2}}\exp
\left(-\frac{B^2}{2k} \right)
\label{dist1}
\end{equation}
where
\begin{equation}
B=\frac{\pi}{3}\sqrt{\frac{\pi}{2}}\,\frac{Jn_0}{\sqrt{M}}
\left\langle\frac{1}{\omega} \right\rangle_0 \equiv
\omega_c\sqrt{M} .
\label{B1}
\end{equation}
Here $n_0$ is the total concentration of HO in the cluster
and $\left\langle 1/\omega\right\rangle_0$ is the $\omega^{-1}$
moment of the normalized initial DOS $g_0(\omega)$. This formula is a
more accurate definition of the characteristic frequency $\omega_c$
introduced in Eq.~(\ref{omc}). Note that the distribution $\rho(k)$
(\ref{dist1}) belongs to an important class of one-sided stable
distributions, Ref.~\cite{WF}.

Due to the combined action of interaction and anharmonicity the DOS
is reconstructed to $\widetilde{g}(\omega) =
2\omega \widetilde{G}(\omega^2)$ with
\begin{eqnarray}
\widetilde{G}(\omega^2) = \left\langle\delta\left( \omega^2 -
\widetilde{\omega}_1^2\right)\right\rangle_{k,\omega_1}
\equiv \nonumber\\
\int\limits_0^{\infty}dk\,\rho(k)\int\limits_0^{\infty}
d\omega_1^2\,G_0(\omega_1^2)\delta\left( \omega^2 -
\widetilde{\omega}_1^2\right)
\label{G1}
\end{eqnarray}
and $G_0(\omega_1^2)\equiv g_0(\omega_1)/2\omega_1$.
Using Eq.~(\ref{1}) and integrating Eq.~(\ref{G1}) we obtain
\begin{eqnarray}
\widetilde{G}(\omega^2) &=& \frac{1}{2}\int\limits_0^\infty
dk\rho\left(k+\frac{1}{2}M\omega^2\right)
G_0\left(\frac{k}{M} \right) + \nonumber\\
&&\int\limits_0^\infty dk\rho(k) G_0\left(\omega^2 +\frac{k}{M}
\right) .
\label{G2}
\end{eqnarray}
For low frequencies, $\omega\ll\omega_c\,$, $\widetilde{G}(\omega^2)
= \mbox{const}$ and $\widetilde{g}(\omega)\propto\omega$, i.e. the
reconstructed DOS is a {\em linear function} of $\omega$. For high
frequencies the first term in Eq.~(\ref{G2}) can be discarded and the
original DOS is reproduced, $\widetilde{G}(\omega^2) =
G_0(\omega^2)$ for $\omega\gg\omega_c$.

\section{Holtsmark method}
\label{sec:holts}

\subsection{Distribution of random forces}
\label{sec:rfd}

Let $x_i$ be a random value with zero mean, $\left<x \right>=0$, and
finite $\left<|x| \right>$ and let ${\bf r}_i$ ($N\to\infty$) be
Poisson-distributed random points in 3-d space with concentration $n_c$.
The distribution function, $P(f)$,  of the random values
\begin{equation}
f=\sum_i\frac{x_i}{r_i^3}
\label{eq:df}
\end{equation}
can then be calculated by the Holtsmark method \cite{H}.
\begin{equation}
P(f) = \left<\delta\left(f-\sum_i\frac{x_i}{r_i^3} \right)
\right> \equiv \frac{1}{2\pi}\int\limits_{-\infty}^{\infty}d\tau
e^{if\tau} F(\tau)
\label{eq:s1}
\end{equation}
with
\begin{equation}
F(\tau) = \left<\exp\left[-i\tau \sum_i\frac{x_i}{r_i^3}\right] \right> ,
\label{eq:s1a}
\end{equation}
where angular brackets denote averaging over $x_i$ and ${\bf r}_i$.
Since the values $x_i/r_i^3$ are independent of each other
\begin{equation}
F(\tau) = \left<e^{-i\tau x/r^3} \right>^N
=\left(1-\frac{1}{V}
\int d^3{\bf r}\left<1-e^{-i\tau x/r^3}
\right>_x \right)^N
\label{eq:s2}
\end{equation}
which in the limit $N\to\infty$ becomes
\begin{equation}
F(\tau) = \exp\left[-n_c\int d^3{\bf r}
\left<1-e^{-i\tau x/r^3} \right>_x
\right] .
\label{eq:t3}
\end{equation}
Using $\left<x\right>=0$ and changing the integration variable to
$y=|\tau| |x| /r^3$ we get
\begin{equation}
F(\tau) = \exp\left[-\frac{4\pi n_c}{3}
|\tau|\left<|x|\right>\int\limits_0^{\infty}
\frac{dy}{y^2}(1-\cos y)\right].
\label{eq:v2}
\end{equation}
The integral equals $\pi/2$ and Eq.~(\ref{eq:s1}) is the Fourier
transform of a Lorentzian distribution of random forces
\begin{equation}
P(f)=\frac{1}{\pi}\frac{\delta f}{f^2+(\delta f)^2}
\label{eq:s3}
\end{equation}
where the width of the distribution is given by
\begin{equation}
\delta f = \frac{2\pi^2 n_c}{3}\left<|x| \right> .
\label{eq:deltaf}
\end{equation}

\subsection{$\rho(k)$ distribution}
\label{sec:rhok}

The same method can be applied to calculate the distribution of
the random quantity $k$
\begin{equation}
k=\sum_i\frac{x_i^2}{r_i^6\omega_i^2} .
\label{eq:r1}
\end{equation}
where the $x_i$ are random and uniformly distributed,
$-x_0/2< x_i<x_0/2$, ${\bf r}_i$ are ($N\to\infty$)
Poisson-distributed random points in 3-d space (concentration $n_0$)
and $\omega_i$ are random frequencies of HO distributed in the interval
\begin{equation}
0<\omega_i<\omega_0
\label{eq:intf}
\end{equation}
with a DOS, $g_0(\omega)$, normalized to unity.

Analogously to (\ref{eq:s1}) the distribution $\rho(k)$  can be
written as
\begin{equation}
\rho(k)=\frac{1}{2\pi}\int\limits_{-\infty}^{\infty}d\tau
e^{ik\tau} K(\tau)
\label{eq:k1}
\end{equation}
with
\begin{equation}
K(\tau) = \left<\exp\left[-i\tau
\sum_i\frac{x_i^2}{r_i^6\omega_i^2}\right] \right>.
\label{eq:ss1}
\end{equation}
Following the steps of the previous subsection we can write
\begin{equation}
K(\tau) = \exp\left[-n_0\int d^3{\bf r}
\left<1-e^{-i\tau x^2/r^6\omega^2} \right>_{x,\omega}
\right]
\label{eq:la3}
\end{equation}
and, introducing the new variable $y=(x^2|\tau|)/(r^6\omega^2)$,
\begin{equation}
K(\tau) = \exp\left[-\frac{2\pi}{3}n_0
\left<\frac{|x|}{\omega}\right>_{x,\omega}\sqrt{|\tau|}\,(\alpha +
i\beta\, {\rm sign}\,\tau) \right]
\label{eq:ba2}
\end{equation}
where
\begin{equation}
\alpha=\int\limits_0^\infty\frac{dy}{y^{3/2}}(1-\cos y),\quad
\beta=\int\limits_0^\infty\frac{dy}{y^{3/2}}\sin y
\label{eq:albe}
\end{equation}
It is straightforward to show that $\alpha=\beta=\sqrt{2\pi}$.
Therefore
\begin{equation}
K(\tau) = \exp\left[-B\sqrt{|\tau|}\,(1+i\,{\rm sign}\,\tau) \right]
\label{eq:kj1}
\end{equation}
with
\begin{equation}
B=\frac{\pi}{3}\sqrt{\frac{\pi}{2}}\,
n_0x_0\left<\frac{1}{\omega}\right>
\label{eq:ma2}
\end{equation}
After integration in (\ref{eq:k1}) with $K(\tau)$ from (\ref{eq:kj1})
we finally get 
\begin{equation}
\rho(k)=\frac{1}{\sqrt{2\pi}}\frac{B}{k^{3/2}}\exp
\left(-\frac{B^2}{2k} \right) .
\label{eq:gh4}
\end{equation}

%%%%%%%%%%%%%%%%%%%%%%%%   REFERENCES   %%%%%%%%%%%%

\end{document}